\documentclass[amsmath,amssymb,aps,prl,twocolumn,longbibliography,nobibnotes]{revtex4-2}

\usepackage{graphicx}
\usepackage{color}
\usepackage{enumitem}

\usepackage{picture}
\usepackage{calc}

\usepackage{subfigure}
\usepackage{dsfont}
\usepackage{color}

\usepackage[utf8x]{inputenc}

\usepackage{tikz}
\usepackage[english]{babel}
\usepackage[T1]{fontenc}
\usepackage{indentfirst}
\usepackage{amsmath}
\usepackage{amssymb}
\usepackage{amsthm}
\usepackage{proof}
\usepackage{eufrak}
\usepackage{graphicx}
\usepackage{psfrag}

\allowhyphens

\newcommand{\ch}{\text{ch}}
\newcommand{\sh}{\text{sh}}

\newcommand{\complex}{\mathbb{C}}

\newcommand{\valos}{\mathbb{R}}

\newcommand{\Li}{{\mathcal{L}}}

\newcommand{\lB}{\mathbb{B}}
\newcommand{\comment}[1]{{\bf\textcolor{green} {comment:}}}

\newcommand{\vev}[1]{\left\langle #1 \right\rangle}
\newcommand{\ket}[1]{{\left|#1\right\rangle}}
\newcommand{\bra}[1]{{\left\langle #1\right|}}

\newcommand{\secc}[1]{\textit{#1}.---}


\usepackage{ifpdf}

\ifpdf
\usepackage{epstopdf}
\usepackage[pdftex,colorlinks,urlcolor=blue,citecolor=blue,linkcolor=blue]{hyperref}
\else
\usepackage[hypertex,colorlinks,urlcolor=blue,citecolor=blue,linkcolor=blue]{hyperref}
\fi
\begin{document}
%\numberwithin{equation}{section}

\title{Constructing Integrable Lindblad Superoperators}

\author{Marius de Leeuw}%
 \email{mdeleeuw@maths.tcd.ie}
 \affiliation{%
School of Mathematics
\& Hamilton Mathematics Institute\\
Trinity College Dublin\\
Dublin, Ireland
}
\author{Chiara Paletta}%
 \email{palettac@maths.tcd.ie}
 \affiliation{%
School of Mathematics
\& Hamilton Mathematics Institute\\
Trinity College Dublin\\
Dublin, Ireland
}
\author{Bal\'azs Pozsgay}%
 \email{pozsgay.balazs@gmail.com}
 \affiliation{%
Department of Theoretical Physics \& \\
MTA-ELTE “Momentum” Integrable Quantum Dynamics Research Group\\
E\"otv\"os Lor\'and University \\
Budapest, Hungary
}

\begin{abstract}
We develop a new method for the  construction of one-dimensional integrable Lindblad
systems, which describe quantum many body models in contact with a
Markovian environment. 
We find several new models with interesting features, such as
annihilation-diffusion processes, a mixture of
coherent and classical particle propagation, and a rectified steady
state current.
We also find new ways to represent known classical integrable
stochastic equations by integrable Lindblad operators.
Our method can be extended to various other situations and
it establishes a structured approach to the study of solvable open quantum systems. 
\end{abstract}

\maketitle

\secc{Introduction}
One-dimensional quantum integrable models
display exotic physical behaviour.
They possess a large number of
conserved charges which constrain the dynamical processes.
As a consequence, isolated integrable
models equilibrate to the Generalized Gibbs Ensemble (GGE)
\cite{rigol-gge,JS-CGGE,rigol-quench-review},
and their large scale transport properties
can be described by Generalized Hydrodynamics (GHD)
\cite{doyon-ghd,jacopo-ghd}. These special properties have been
investigated in a number of experiments with cold atoms, see for
example \cite{QNewtonCradle,gge-experiment1,ghd-experimental-atomchip,LL-GHD-exp}.
However, in experiments
there are always integrability breaking interactions within the system, and also between the
system and its environment.
Recently a number of works were devoted to the question
whether the integrability breaking effects can be handled within the
GGE and the GHD, see for example \cite{zala-tGGE,vasseur-breaking,integr-breaking-potential,space-time-inhom,alvise-js-adiabatic-formation,doyon-weak-breaking,Lindblad-noise,jerome-atom-loss}.

The interactions with the environment can be described by the Lindblad equation if the response of the
environment is Markovian \cite{lindblad-eredeti,lindblad-intro,rivas2010markovian}.
While most Lindbladians explicitly break integrability, 
there are cases when the Lindblad equation itself is integrable, or it shows certain features of integrability.
Finding solvable examples is important because then the
non-equilibrium steady states (NESS) and the relaxation towards them can be studied with exact
methods. Exact solutions could be used to justify the approximations involving time
dependent GGE or GHD \cite{zala-tGGE,Lindblad-noise,jerome-atom-loss,bose-hubbard-loss}, and 
they are also relevant for quantum circuits \cite{lindblad-circuit}. For potential experimental
applications see \cite{lindblad-exp0,lindblad-experiments,zala-trapped,bose-hubbard-loss,nava-rossi-giuliano}.

Lindblad systems showing various types of solvability include
{\it a)} models solvable by free fermion techniques \cite{third-quantization,third-quantization-2,katsura-lindblad-1,eric-lindblad}
{\it b)} boundary driven spin chains
that allow for the construction of their NESS \cite{prosen-boundary-lindblad-1,prosen-boundary-lindblad-2,boundary-lindblad-mps,prosen-exterior-lindblad,enej-thesis,spin-helix-boundary},
{\it c)} triangular Lindblad superoperators that allow for the computation of the Liouvillian spectrum
\cite{marko-lindblad-1,japanok-lindblad}, {\it d)} models where the integrability is established separately
for different subspaces of the full operator space \cite{essler-piroli-lindblad};
and {\it e)} Yang-Baxter integrable Lindblad systems {with local jump operators in the bulk}
\cite{essler-prosen-lindblad,katsura-lindblad-2,essler-lindblad-review}.  
The latter type of models are related to solutions of the Yang-Baxter equation, which is a central
algebraic relation in the theory of integrable models \cite{Korepin-Book}.
So far there has been no structured approach to Yang-Baxter integrable
Lindblad systems: the examples in the literature were found through relations to known spin ladder systems
\cite{essler-prosen-lindblad,katsura-lindblad-2,essler-lindblad-review},
but there was no method for actually constructing new integrable
cases.

In the present paper we fill this gap and initiate a systematic
classification of Yang-Baxter integrable Lindblad systems. We explore
certain sections of the parameter space for such models 
and we find several new cases. Our main
method is borrowed from  
\cite{marius-classification-1,marius-classification-2,marius-classification-3,marius-classification-4},
where solutions of the Yang-Baxter equation were classified in various other circumstances.
The new Lindblad systems correspond to novel solutions of the
Yang-Baxter equation and exhibit interesting features, such as
a tunable coupling to the environment and
a mixture of coherent and classical transport.

\secc{The Lindblad equation}
We consider a spin-1/2 chain of length $L$, with 
a Hamiltonian $ H=\sum_{j=1}^L h_{j,j+1}$, where $h_{j,j+1}$ is the Hamiltonian density acting on the
neighboring sites $j$ and $j+1$.
We assume periodic boundary conditions. 
We also assume that our model is in contact with a thermal
environment that is Markovian, so that its internal dynamics evolves
much faster than that of our model.
In this case, the time dependence of the density matrix of
our model is well approximated by the Lindblad master equation \cite{lindblad-eredeti,lindblad-intro,manzano2018harnessing,breuer2002theory}:
\begin{equation}
  \label{lindblad}
  \dot \rho\equiv\Li\rho=i[\rho, H]+\sum_a \Big[
     \ell_a\rho  \ell_a^\dagger-\frac{1}{2}\{ \ell^\dagger_a  \ell_a,\rho\}\Big].
\end{equation}
Here $\Li$ is the so-called
Lindblad superoperator.
In \eqref{lindblad} the commutator term describes coherent time evolution dictated by the
Hamiltonian, and the $\ell_a$ are the so-called Lindblad or jump
operators, which describe dissipative processes mediated by the
environment.
The summation above runs over all
 interactions with the environment; we focus on local
 processes in the bulk
 with one family of local jump operators
$\ell_a\equiv \ell_{j,j+1}$, which act on sites $j$ and $j+1$ of the spin-chain. 

\secc{Classical flows}
Before discussing the integrable cases we explain an important
connection to integrable classical stochastic equations, which have a rich history themselves
\cite{stochastic-book}. It turns out that the Lindblad equation is capable of realizing the classical
flows on the diagonal of the density matrix.
The diagonal elements of $\rho$ can be understood as classical probabilities of
finding the system in the given state, and in some models the operator space spanned by the diagonal elements
is kept invariant by the Lindblad superoperator. In such cases the flow equation \eqref{lindblad} for
$\rho$ can be projected to the diagonal elements  $\bra{n_1,\dots,n_L}\rho
\ket{n_1,\dots,n_L}\equiv P(n_1,\dots,n_L)$, where  $\ket{n_1,n_2,\dots,n_L}$, $n_j=\uparrow,\downarrow$ are the
vectors of the computational basis. 
Defining the vector
\begin{equation}
  \ket{P}=P(n_1,\dots,n_L) \ket{n_1,\dots,n_L},
\end{equation}
we obtain a flow equation 
$\partial_t \ket{P}=   W \ket{P}$,
where $ W=\sum_{j=1}^L w_{j,j+1}$ is the generator of the classical flow,
with its matrix elements given by the corresponding projection of $\Li$. Lindblad equations that
keep the diagonal of $\rho$ invariant will be called {\it diagonal
  preserving}.

If the initial $\rho$ is diagonal in such a model, then
it will stay diagonal and the classical flow will be
realized. However, preservation of the  diagonal does not imply that the orthogonal complement of
the diagonal subspace is also conserved, so  for generic configurations we can still expect quantum
effects in the time evolution. 

An example for a diagonal conserving model was discussed in \cite{stoch-coh-mixt} (see also
\cite{eisler-lindblad-xx,essler-piroli-lindblad}).
Here $h_{j,j+1}=0$ and the model has two families of jump operators 
\begin{equation}
  \label{noti}
   \ell^R_{j,j+1}=\sqrt{\varphi_R} \sigma_j^-\sigma_{j+1}^+,\qquad
    \ell^L_{j,j+1}=\sqrt{\varphi_L} \sigma_j^+\sigma_{j+1}^-,
\end{equation}
with $\varphi_{L,R}\ge 0$.
The resulting classical flow was found to be
the 
Asymmetric Simple Exclusion Process (ASEP), with the generator being
\begin{equation}
  \begin{split}
  w_{j,j+1}=\varphi_R& [\sigma_j^-\sigma_{j+1}^+-  n_j(1-n_{j+1})]+\\
 &+\varphi_L [\sigma_j^+\sigma_{j+1}^--  (1-n_j)n_{j+1}].
  \end{split}
\end{equation}
It is known that the ASEP is related to the Heisenberg XXZ spin chain
\cite{asep-xxz} and thus it is integrable.
However, the integrability properties of the
superoperator itself were not investigated in \cite{stoch-coh-mixt,essler-piroli-lindblad}. We found
that the Lindblad system given by \eqref{noti} is not Yang-Baxter
integrable, but we found
integrable ``parent'' Lindbladians in some special cases, see below.

\secc{Lindblad equation and spin ladders} We treat the
Lindblad superoperator as a (non-Hermitian) Hamiltonian of a spin ladder, where
the two legs correspond to the ``bra'' and ``ket'' sides of the
density matrix \cite{essler-prosen-lindblad}.
We use the computational basis to identify the Hilbert
space $\mathcal{H}$ with its dual $\mathcal{H^*}$. Then the density
matrix 
$\rho \, \in \,  \mathcal{H}\otimes \mathcal{H}^*$
can be represented by an element of the tensor product
$\mathcal{H}^{(1)}\otimes\mathcal{H}^{(2)}$
which in turn is interpreted as the
Hilbert space of a spin ladder. 

As mentioned, we consider one family of jump operators in the bulk.
Then the superoperator can be written in the spin ladder representation as
$\Li=\sum_j \Li_{j,j+1}$, with
\begin{align}
  \label{bulkli}
  \Li_{j,j+1}=&~-i h^{(1)}_{j,j+1}+i  h^{(2)*}_{j,j+1}+  \ell^{(1)}_{j,j+1}  \ell^{(2)*}_{j,j+1} \nonumber \\
&~  -\frac{1}{2} \ell^{(1)\dagger}_{j,j+1}  \ell^{(1)}_{j,j+1}-\frac{1}{2}  \ell^{(2)T}_{j,j+1}  \ell^{(2)*}_{j,j+1}.
\end{align}
Above the
superscript $^T$ denotes transpose and the
asterisk stands for complex conjugation component wise. 
For any operator $A$ the notation $A^{(1)}$ or $A^{(2)}$ means that it acts only on $\mathcal{H}^{(1)}$ or
$\mathcal{H}^{(2)}$.
It is our goal to find integrable spin ladders with a (non-Hermitian) Hamiltonian having the structure \eqref{bulkli}.

\secc{Yang-Baxter integrability}
A large class of quantum integrable models can be constructed from solutions of the celebrated
Yang-Baxter equation (YBE) \cite{Korepin-Book}.
The YBE concerns the so-called $R$-matrix $R(u,v)$, which acts on $V\otimes V\simeq \complex^d\otimes\complex^d$, and $u$ and $v$ are the so-called spectral parameters.
The YBE reads
\begin{multline}
  \label{eq:YBE}
  R_{12}(u_1,u_2) R_{13}(u_1,u_3)R_{23}(u_2,u_3)=\\
  =R_{23}(u_2,u_3)R_{13}(u_1,u_3)R_{12}(u_1,u_2),
\end{multline}
which is an equality for operators acting on $V\otimes V\otimes V$,
and each spectral parameter is associated to a space $V$.
The solutions generate
integrable spin chains;
for the details of the construction we refer to 
\cite{Korepin-Book,marius-classification-1}, and here we just review a few key statements.

If the regularity condition $R(u,u)=P$ holds with
$P$ being the permutation operator, then we can define a family of Hamiltonian densities
as
\begin{equation}
  \label{hR}
  h(u)=\left. P \partial_v R(v,u)\right|_{v=u}.
\end{equation}
Each $h(u)$ describes a nearest neighbor interacting integrable
chain with local spaces $\simeq\complex^d$. The models are generally different for different values of
$u$. However, if the $R$-matrix is of 
difference form $R(v,u)=R(v-u)$, then the $u$-dependence drops out.

The models possess a commuting set of charges $Q_\alpha$ (with range $\alpha$), such that
the Hamiltonian is a member of the series.
Identifying $Q_2=H$, the higher charges can be obtained either from the
transfer matrix \cite{Korepin-Book}, or  they can be generated
 with the so-called boost operator $\lB$
\cite{Tetelman, HbbBoost,Loebbert:2016cdm} as
\begin{align}
  \label{boostop}
&Q_{\alpha+1}(u) \sim [\lB,Q_\alpha(u)],
&& \alpha>1.
\end{align}
The boost operator is a differential operator and it depends on the
Hamiltonian density:
\begin{equation}\label{eq:boost}
\lB:=\partial_u +  \sum_{j=-\infty}^\infty j\, h_{j,j+1} (u).
\end{equation}
{This formal infinite sum gives a well  defined commutator in \eqref{boostop}.}

\secc{Construction of new models}
Let us review the method of
\cite{marius-classification-1,marius-classification-2,marius-classification-3,marius-classification-4}
to find new integrable Hamiltonians; the
extension to Lindblad systems is discussed afterwards. The key
idea is to start with a Hamiltonian density $h_{j,j+1}(u)$,
then find the first non-trivial higher charge $Q_3(u)$ using
\eqref{boostop} and
\eqref{eq:boost}, and to  impose the condition $[Q_2(u),Q_3(u)] =
0$. This leads to a set of
coupled first order, non-linear, differential equations for the
components of $h(u)$, which can be solved if an initial Ansatz  for
$h(u)$ is given. A priori it is not clear whether the commutativity between $Q_2(u)$ and
$Q_3(u)$ is enough to ensure the existence of the remaining higher
charges, but in all previous cases this was found to be
sufficient. The $R$-matrix is afterwards found by solving the
so-called Sutherland-equation, taking into account \eqref{hR} {and the regularity condition}.
This strategy has proven to be
successful in several cases and  led to many new solutions of the Yang-Baxter equation \cite{marius-classification-1,marius-classification-2,marius-classification-3,marius-classification-4}. 

\secc{Integrable Lindblad superoperators}
For Lindblad systems the key is to find a
(non-Hermitian) Hamiltonian density for a spin ladder, which takes
the special Lindblad form \eqref{bulkli}. However, 
now we also allow for an $u$-dependence of the matrices $h$ and $\ell$, where $u$ is the spectral parameter used
in the $R$-matrix.
Different values of $u$ will generally give different physical
systems, except if the corresponding $R$-matrix is of difference form.
We substitute this Ansatz into our integrability constraint and then solve the corresponding
set of differential equations for the components of the operators
$h(u)$ and $\ell(u)$.

Strictly speaking we are imposing too strong a
condition: 
we require that  we find a Lindblad
superoperator for \textit{any} value of $u$, whereas
it would be enough to find a solution
of the Yang-Baxter equation such that the decomposition
\eqref{bulkli} works for \textit{some} value of
$u$. However, this strong requirement leads to a
structured approach and it allows for a number of non-trivial
Lindblad systems.

Once a model is found, we can apply simple transformations which lead to seemingly different
matrices, nevertheless describing the same physical behaviour. These
transformations include {\it a)}
unitary basis transformations, {\it b)} space- and spin-reflection, {\it c)} trivial shifts $h\mapsto h+r$ with $r\in\valos$,
{\it d)} telescopic operator shifts $h_{j,j+1}\mapsto h_{j,j+1}+o_j-o_{j+1}$, where $o_{j}$ is a one-site
operator, eventually adding up to zero on the periodic chain, {\it e)}
combined shifts
\begin{align}
  \label{hladd}
 (h,\ell)  \mapsto (h+i (z \ell^\dagger -z^* \ell), \ell+2z),\quad z\in\complex,
 \end{align}
 which leave the superoperator invariant, and {\it f)} a combined re-scaling
\begin{align}
  \label{hlphase}
(h,\ell) \mapsto (r h, \sqrt{r} e^{i\mu}\ell),\quad r\in\valos^+,\mu\in\valos,
\end{align}
which only affects the choice of the unit of time.
Below we always choose a representation such that the matrices of either $h$ or $\ell$ are as simple as possible.

\secc{Partial classification}
We carve out a section of the space of integrable superoperators by restricting to
specific forms of $h$ and $\ell$. The two choices we investigate are the
following: {\it A)} Lower triangular $\ell$ operators with at most two elements below the
diagonal. 
 {\it B)}  $h$ and $\ell$ operators
 that both conserve the total $S^z$ quantum number.

Curiously both restrictions  {\it A)} and  {\it B)} only
allow for very specific Hamiltonian densities: We found that either
$h$ is diagonal, or it is given by
\begin{equation}
  \label{HXX}
  h_{j,j+1}=\frac{1}{2}\left[e^{i\phi}\sigma_j^+\sigma_{j+1}^-+e^{-i\phi}\sigma_j^-\sigma_{j+1}^+\right],\quad \phi\in\valos.
\end{equation}
This Hamiltonian density describes a free fermionic hopping model; the
factor of $1/2$ is added only for later convenience.
The angle $\phi$ can be understood as a homogeneous twist along the
chain, and the $\phi=0$ point corresponds to the XX spin
chain. Alternatively, the model can be interpreted as the XX chain perturbed by a
Dzyaloshinskii–Moriya interaction term, \cite{Dzyaloshinsky-Moriya-xxz}.

We  use the particle picture to interpret our models. We
choose the vacuum state as the state with all spins down, and an up spin
is interpreted as a particle;
the operators
$n_j=(1+\sigma^z_j)/2$ measure the local
occupation numbers.

We now give a list of integrable Lindbladians, almost
all of which are new \footnote{We omit some models with less interesting physical
properties, such as those with diagonal $h$ and $\ell$ operators.  A complete list will be presented
elsewhere.}. The $R$-matrices of the models are presented for all cases in \footnote{Supplemental
Materials to ``Constructing Integrable Lindblad Superoperators''.}.

\secc{Model A1} If we require that $\ell$ is lower triangular with just one
element below the diagonal then we find
only one solution. It can be represented with $h$ given by \eqref{HXX} and the Lindblad
operator being
\begin{equation}
  \ell_{j,j+1}=\sigma_j^-\sigma_{j+1}^+-i e^{i \phi}(1-n_j)n_{j+1}.
  \label{lmodea1}
\end{equation}
The corresponding $R$-matrix is of difference form and seems to be new.
Direct computation shows that the model is diagonal preserving,
 and for the diagonal flow we find the generator
\begin{equation}
  \label{wtasep}
  w^{\text{TASEP}}_{j,j+1}
  =
\sigma_j^-\sigma_{j+1}^+- n_j (1-n_{j+1}).
\end{equation}
This is the generator of the Totally Asymmetric Simple Exclusion Process (TASEP)
\cite{stochastic-book}. To our best knowledge this is the first realization of the TASEP using an
integrable Lindbladian.
%\green{To explain parent, what if we say: To our best knowledge this is the first realization of the TASEP starting from a "parent"
%integrable Lindbladian.}

\secc{Model A2} If we require a lower triangular $\ell$ with two
elements below the diagonal and a non-diagonal $h$ we again find
only one solution, which has an $R$-matrix of difference form. The Hamiltonian density is given by
\eqref{HXX} with $\phi=\tau \pi/2$ and the Lindblad operator is now
\begin{equation}
\ell_{j,j+1}=n_{j+1} +\tau \sigma_j^-\sigma_{j+1}^++\sigma_j^-\sigma_{j+1}^-,
\end{equation}
where $\tau=\pm1$.
This jump operator describes particle propagation to the right
and local two-body loss. The model is diagonal preserving with the
generator being
\begin{equation}
w_{j,j+1}=-n_j+\sigma_j^-\sigma_{j+1}^++\sigma_j^-\sigma_{j+1}^-.
\end{equation}
This corresponds to the totally asymmetric limit of the
diffusion-annihilation model treated in
\cite{lushnikov-annihilation,lushnikov-annihilation2,asymmetric-two-body-anni-free-fermion}.
To our best knowledge this is the first time that this generator is
embedded into an integrable Lindbladian.

Let us now turn to the models where both $h$ and $\ell$ conserve the
global $S^z$ operator.

\secc{Model B1} In this case $h=0$ and the Lindblad operator is
\begin{equation}
  \ell=
  \begin{pmatrix}
    \tau & 0 & 0 & 0  \\
    0 & 0 & 1 & 0 \\
    0 & 1 & 0 & 0 \\
    0 & 0 & 0 & \kappa 
  \end{pmatrix},
\end{equation}
where $\tau=\pm 1$, $\kappa=\pm 1$. In the case of $\tau=\kappa=1$ the superoperator is
equivalent to the Hamiltonian of the $SU(4)$-invariant chain; this
case was listed in \cite{essler-lindblad-review}. The $R$-matrix is of
  difference form for all four choices, and for $\tau\kappa\ne 1$ it seems to be new.
The model is diagonal preserving in all four cases  and the
generator is
\begin{equation}
  w_{j,j+1}=
  \sigma_j^+\sigma_{j+1}^-+
  \sigma_j^-\sigma_{j+1}^+-
  n_j-n_{j+1}+2n_j n_{j+1}.
\end{equation}
This is the generator of the Symmetric Simple Exclusion Process (SSEP)
\cite{stochastic-book}. To our best knowledge it is the first
time that the SSEP is realized by an integrable Lindbladian.
The parameters $\tau,\kappa$ have an effect on the off-diagonal
sectors of the superoperator, and they influence the Liouvillian spectrum.

\secc{Model B2} The Hamiltonian is given by
\eqref{HXX} and the Lindblad operator is given by
 \begin{equation}
   \frac{\ell(u)}{\beta(u)}=
   \begin{pmatrix}
    \ch(u) & 0 & 0 &0 \\
 0   & 1 & i \sh(u) e^{i \phi } &0 \\
 0   & -i \sh(u) e^{-i \phi } & -1 & 0\\
 0   & 0& 0& -\ch(u)
    \end{pmatrix}
  \end{equation}
with $\beta(u)= (\gamma/(2\, \gamma \, \ch (2 u)+2))^{1/4}$
and $\gamma\geq 0$ being a fixed coupling constant.
For $u=\phi=0$ this system is equivalent to the XX
model with dephasing  noise 
treated in \cite{essler-prosen-lindblad}, which in turn corresponds to
the Hubbard model with imaginary coupling. The $R$-matrix is known and it is not of difference
form \cite{Hubbard-Book}.

For $u\ne 0$ the model can be 
interpreted as the inhomogeneous version of the Hubbard model
\cite{essler-lindblad-review,Murakami-Hubbard-inhom}, and it involves
a mixture of coherent and stochastic particle propagation.
The particle current $J_{k}$ can be found from the continuity
relation $dn_k/dt=J_{k-1}-J_{k}$ and the Lindblad equation, and it is
given by
\begin{equation}
  J_k=(1-2\beta^2(u) \sh(u)) J_k^{0}
  +\beta^2(u)\sh^2(u)(n_k-n_{k+1}).
\end{equation}
Here
\begin{equation}
  \label{cohJ}
  J_k^0= \frac{i}{2} (e^{i\phi}\sigma_k^+\sigma_{k+1}^--e^{-i\phi}\sigma_k^-\sigma_{k+1}^+) ,
\end{equation}
is the current of the coherent time
evolution dictated by \eqref{HXX}, and the remaining terms describe stochastic transport.

\secc{Model B3} This is a completely new model. The Hamiltonian is
given again by \eqref{HXX} and 
\begin{align}
&{\ell}={\sqrt{\frac{\gamma}{2}}}\left(
\begin{array}{cccc}
 \gamma & 0 & 0 & 0 \\
 0 & 1 & i(\gamma-1) e^{i \phi } & 0 \\
 0 & -i(\gamma+1) e^{-i \phi } & -1 & 0 \\
 0 & 0 & 0 & \gamma \\
\end{array}
\right).
\end{align}
Here $\gamma\ge 0$ is a
coupling constant. The $R$-matrix is of difference form and it appears to be new.
At the special point $\gamma=1$ the model is diagonal preserving
and it describes the TASEP.
For generic values of $\gamma$ the model
describes a mixture a quantum and classical transport.
The particle current is now given by
\begin{equation}
  \begin{split}
    J_{k}&=
(1-\gamma^2)J_k^0+
 \frac{\gamma(1+\gamma)^2}{2} n_k(1-n_{k+1})- \\
  &  \hspace{1cm}- \frac{\gamma(1-\gamma)^2}{2} (1-n_k) n_{k+1}
    \\
  \end{split}
\end{equation}
with $J_k^0$ given by \eqref{cohJ}. The stochastic terms explicitly
break space reflection symmetry: they describe the current in the
ASEP. 

\secc{Steady states} The dynamics of the models
lead to the formation of non-equilibrium steady-states $\rho$, which
satisfy $\Li\rho=0$. Here we focus on Models B2 and B3.

In Model B2 the NESS is
the infinite temperature state \cite{essler-prosen-lindblad}.
In contrast, in Model B3 we find that the NESS is a current carrying
mixed state, which actually degenerates into a pure state if the
compatibility condition $e^{i (\phi+\pi/2)L}=1$ holds. For the XX model 
($\phi=0$) this means $L\equiv 0\mod{4}$. In such a case let us
consider $\rho_h=\ket{\Psi}\bra{\Psi}$ where $\ket{\Psi}$ is a spin
helix state given by
\begin{equation}
  \ket{\Psi}=\otimes_{j=1}^L
\frac{1}{\sqrt{2}}  \begin{pmatrix}
1 \\ e^{i j(\phi+\pi/2)}
  \end{pmatrix}.
\end{equation}
Direct computation shows that $H\ket{\Psi}=0$ and $\Li
\rho_h=0$ for every $\gamma$. The superoperator conserves particle
number, thus the NESS' in the sectors with fixed spin are given by
the appropriate projections of $\rho_h$. Numerical studies on small systems show that these are the
unique steady states in the various spin sectors.

The particle current in the projected states can be
computed easily, and in the thermodynamic limit we
find
\begin{equation}
\lim_{L\to\infty}  \vev{J_k}=(1+\gamma^2) \vev{n}\big(1-\vev{n}\big).
\end{equation}
We conjecture that this formula holds generally, even if the
compatibility condition is not met.

We observe a remarkable phenomenon: there is a finite particle current even if the
Lindblad coupling $\gamma$ is tuned back to zero.
This phenomenon is understood as a
pumping effect \cite{pumping,rectification}.
The jump operators are coupled to the
coherent current, thus they build up its mean value over time.
The current itself is conserved, thus it cannot decay.
Eventually a current carrying state is produced no matter how small the coupling is. Our
model presents an exactly solvable case for this phenomenon, earlier
discussed in \cite{pumping,rectification}. For the same effect with
boundary driving see \cite{spin-helix-boundary}. 

\secc{Discussion}
We presented a method for the construction of
integrable Lindblad systems, and found new models for spin-1/2 chains.
As a by-product we also discovered apparently new solutions of the
Yang-Baxter equation, which can also describe Hermitian spin ladders
for different choices of their parameters.

  In some cases our integrable Lindblad
superoperators support known classical flows on the diagonal of the
density matrix.  Similar embeddings
were already known in the literature 
\cite{stoch-coh-mixt,eisler-lindblad-xx,essler-piroli-lindblad}, but the existence of 
Yang-Baxter integrable superoperators behind the classical flows
is a new result.

It would be important to continue the classification
commenced in this work; so far we  only explored a limited
parameter space for the models.
Our methods could potentially lead to further integrable
Lindblad systems, including models with particle creation and
annihilation processes, multiple families of jump
operators, higher dimensional local spaces, and models with boundaries.
Furthermore, it is desirable to find the Bethe Ansatz solution to
Model B3, which would lead to an understanding of its relaxation
dynamics.

\begin{acknowledgments}
\paragraph{Acknowledgments.}
%%%%%%%%%%%%%%%%%%%%%%%%%%%%%%%%%%%%%%%%%%%%%%%%%%%%%%%%%%%%%%%%%%%%%%%%%%%%%%%%

We would like to thank 
Zala Lenar\v{c}i\v{c}, 
Andrea Nava,
Lorenzo Piroli, 
Toma\v{z} Prosen,
 Florentin Reiter,  
 Ana L. Retore and
 Marco Rossi for useful
discussions. M.dL. was supported by SFI, the Royal Society and the EPSRC
for funding under grants UF160578,
RGF$\backslash$EA$\backslash$181011,
RGF$\backslash$EA$\backslash$180167 and 18/EPSRC/3590.  
C.P. is supported by the grant RGF$\backslash$EA$\backslash$181011. 

\end{acknowledgments}

%\bibliographystyle{utphys}
%\bibliography{lindblad-paper}
\providecommand{\href}[2]{#2}\begingroup\raggedright\endgroup

\widetext
\newpage

\setcounter{equation}{0}
\setcounter{figure}{0}
\setcounter{table}{0}
\setcounter{page}{1}
\makeatletter
\renewcommand{\theequation}{S\arabic{equation}}
\renewcommand{\thefigure}{S\arabic{figure}}
\renewcommand{\bibnumfmt}[1]{[S#1]}
\renewcommand{\citenumfont}[1]{S#1}

\begin{center}
\textbf{Supplemental Material:}\\
\smallskip
\textbf{Constructing Integrable Lindblad Superoperators}
\end{center}

\setcounter{equation}{0}
\setcounter{figure}{0}
\setcounter{table}{0}
\setcounter{page}{1}
\makeatletter
\renewcommand{\theequation}{S\arabic{equation}}
\renewcommand{\thefigure}{S\arabic{figure}}
\renewcommand{\bibnumfmt}[1]{[S#1]}
\renewcommand{\citenumfont}[1]{S#1}

\section{Notations}

Below we present a list of the integrable Lindbladians treated in the
main text. For each model we give here the matrix representation of
the Hamiltonian density $h$ and the jump operator $\ell$.
We also present the corresponding $R$-matrix, solution of the Yang-Baxter equation
\begin{align}
  R_{12}(u_1,u_2) R_{13}(u_1,u_3)R_{23}(u_2,u_3)
  =R_{23}(u_2,u_3)R_{13}(u_1,u_3)R_{12}(u_1,u_2).
\end{align}

We apply the following conventions. The components of a single spin
are denoted as
\begin{equation}
\left(  \begin{array}{c}
 \psi_{\uparrow} \\
\psi_{\downarrow} \\
\end{array}\right).
\end{equation}

For the tensor product of two spins we apply the notation
\begin{align}
  \label{twoseg}
\left(
\begin{array}{c}
 \psi_{\uparrow} \\
\psi_{\downarrow} \\
\end{array}
\right)\otimes \left(
\begin{array}{c}
 \psi_{\uparrow} \\
 \psi_{\downarrow} \\
\end{array}
\right)\equiv \left(
\begin{array}{c}
 \psi_{\uparrow\uparrow} \\
 \psi_{\uparrow\downarrow} \\
 \psi_{\downarrow\uparrow} \\
 \psi_{\downarrow\downarrow} \\
\end{array}
\right).
\end{align}

The density matrix of a two-site segment is given by
\begin{equation}
  \rho_{ab}^{cd}=\psi_{ab} \psi^*_{cd},\qquad a,b,c,d=\uparrow,\downarrow.
\end{equation}

In the superoperator formalism we treat such a density matrix as a
vector with $16$
components. The identification of the components is given by
\begin{align}
  \label{rhoindex}
\rho=\left(
\begin{array}{cccc}
 \rho _{\uparrow  \uparrow}^{\uparrow \uparrow} & \rho _{\uparrow  \uparrow}^{\uparrow    \downarrow}
  & \rho _{\uparrow \uparrow}^{\downarrow \uparrow} & \rho _{\uparrow \uparrow}^{\downarrow \downarrow} \\[1ex]
 \rho _{\uparrow \downarrow}^{\uparrow \uparrow} & \rho _{\uparrow    \downarrow}^{\uparrow \downarrow}
  & \rho_{\uparrow\downarrow}^{\downarrow  \uparrow}    & \rho _{\uparrow \downarrow}^{\downarrow \downarrow} \\[1ex]
 \rho _{\downarrow \uparrow}^{\uparrow \uparrow} & \rho _{\downarrow \uparrow}^{\uparrow \downarrow} & \rho _{\downarrow \uparrow}^{ \downarrow \uparrow} & \rho _{\downarrow \uparrow}^{ \downarrow \downarrow} \\[1ex]
  \rho_{\downarrow \downarrow}^{ \uparrow \uparrow} & \rho_{\downarrow    \downarrow}^{  \uparrow \downarrow}
  & \rho _{\downarrow \downarrow}^{ \downarrow \uparrow} & \rho _{\downarrow \downarrow}^{ \downarrow \downarrow} \\
\end{array}
\right)=
  \begin{pmatrix}
    \rho_1 &  \rho_2 &  \rho_3 &  \rho_4  \\
    \rho_5 &  \rho_6 &  \rho_7 &  \rho_8  \\
    \rho_9 &  \rho_{10} &  \rho_{11} &  \rho_{12}  \\
     \rho_{13} &  \rho_{14} &  \rho_{15} &  \rho_{16}  \\
  \end{pmatrix}.
\end{align}

The $4\times 4$ matrices $h(u)$ and $\ell(u)$ act on the vectors of the two-site
segment as given by \eqref{twoseg}.
The superoperator $\Li(u)$ can be
represented by a matrix of size $16\times 16$ and it acts on the
components of $\rho$ as indexed by \eqref{rhoindex}.

{For the $R$-matrix we will denote by $R_i^j$ the element corresponding to $j$-row and $i$-column. 
The $R$-matrix satisfies the conditions
\begin{align}
&R(u,u)=P,
&  \Li(u)=\left. P \partial_v R(v,u)\right|_{v=u},
  \label{hR2}
\end{align}
}
with $P$ being the permutation operator acting on two copies of $\complex^4$, and 
\begin{align}
  \label{bulkli2}
  \Li(u)=&~-i h^{(1)}(u)+i  h^{(2)*}(u)+  \ell^{(1)}(u)  \ell^{(2)*}(u)  -\frac{1}{2} \ell^{(1)\dagger}(u)  \ell^{(1)}(u)-\frac{1}{2}  \ell^{(2)T}(u)  \ell^{(2)*}(u).
\end{align}
Here $^T$ denotes transpose and the asterisk stands for complex conjugation element wise.

If the $R$-matrix is of difference form, that is $R(v,u)=R(v-u)$, then the superoperator does not
depend on $u$ and we will suppress it in the notation. 

We choose the multiplicative normalization of the $R$-matrix such that
\eqref{hR2} are
satisfied.

In all the models, $\phi$ is a real constant.

\section{Model A1}

We have
\begin{align}
&h=\frac{1}{2}\left(
\begin{array}{cccc}
 0 & 0 & 0 & 0 \\
 0 & 0 & e^{i \phi } & 0 \\
 0 & e^{-i \phi } & 0 & 0 \\
 0 & 0 & 0 & 0 \\
\end{array}
\right),
&&\ell=\left(
\begin{array}{cccc}
 0 & 0 & 0 & 0 \\
 0 & 0 & 0 & 0 \\
 0 & 1 & -i e^{i \phi } & 0 \\
 0 & 0 & 0 & 0 \\
\end{array}
\right)
\end{align}

The non-zero entries of the $R$-matrix are
\begin{align}
\nonumber &R_1^1=R^4_{13}=R^6_6=R^{11}_{11}=R^{16}_{16}=1,\\
\nonumber & R^2_5=R^5_2=R^3_9=R^9_3=R^8_{14}=R^{14}_8=R^{12}_{15}=R^{15}_{12}=e^{-\frac{u}{2}},\\
\nonumber &-\frac{i R^2_2}{e^{i \phi }}=i e^{i \phi } R^3_3 =R^4_4=i e^{i \phi }R^8_8 =-\frac{i R^{12}_{12}}{e^{i \phi }}=2 e^{-\frac{u}{2}} \sinh \left(\frac{u}{2}\right),\\
\nonumber & -e^{i \phi }R^7_4=\frac{R^{10}_4}{e^{i \phi }}= 2 i e^{-u} \sinh \left(\frac{u}{2}\right),\\
\nonumber & R^7_{10}=R^{10}_7=R^{13}_4=e^{-u}.
\end{align}

An alternative representation for the same super-operator can be
obtained by an additive shift to $\ell$, leading to
  \begin{align}
  &   h=0,\qquad
&&  \ell=i e^{i \phi }\left(
\begin{array}{cccc}
 1 & 0 & 0 & 0 \\
 0 & 1 & 0 & 0 \\
 0 & -i e^{-i \phi } & 0 & 0 \\
 0 & 0 & 0 & 1 \\
\end{array}
\right).
     \end{align}

\section{Model A2}

We have
\begin{align}
&h=\frac{\tau }{2}\left(
\begin{array}{cccc}
 0 & 0 & 0 & 0 \\
 0 & 0 & i & 0 \\
 0 & -i  & 0 & 0 \\
 0 & 0 & 0 & 0 \\
\end{array}
\right),
&&\ell=\left(
\begin{array}{cccc}
 1 & 0 & 0 & 0 \\
 0 & 0 & 0 & 0 \\
 0 & \tau  & 1 & 0 \\
 1 & 0 & 0 & 0 \\
\end{array}
\right),
\end{align}
with $\tau=\pm1$.

The non-zero entries of the $R$-matrix are
\begin{align}
\nonumber &R^{16}_{16}= R^4_{13}= 1,\\
\nonumber & \tau R^8_3= \tau R^{12}_2  =-\tau R^7_4= -\tau R^{10}_4 = R^6_1= R^{11}_1= 2 e^{-u} \sinh \left(\frac{u}{2}\right),\\
\nonumber &-\tau  R^{12}_{12} = -\tau R^8_8 =R^4_4= R^8_9= R^{12}_5= R^{16}_1= 1-e^{-u},\\
\nonumber &R^2_5= R^3_9= R^8_{14}= R^{14}_8= R^{12}_{15}= R^{15}_{12}= e^{-\frac{u}{2}},\\
\nonumber &R^9_3= R^5_2= e^{-\frac{3 u}{2}},\\
&R_1^1= R^6_6= R^{11}_{11}= R^7_{10}= R^{10}_7= R^{13}_4= e^{-u}.
\end{align}

\section{Model B1}

We have
\begin{align}
&h=0,
&&\ell=\left(
\begin{array}{cccc}
 \tau  & 0 & 0 & 0 \\
 0 & 0 & 1 & 0 \\
 0 & 1 & 0 & 0 \\
 0 & 0 & 0 & \kappa  \\
\end{array}
\right),
\end{align}
$\tau=\pm1$ and $\kappa=\pm1.$

The non-zero entries of the $R$-matrix are
\begin{align}
\nonumber &R_1^1= R^{16}_{16}= e^{-u} (u+1),\\
\nonumber&R^6_6= R^{11}_{11}= e^{-u} (\kappa  \tau  u+1) ,\\
\nonumber &R^2_5= R^5_2= R^3_9= R^9_3= R^4_{13}= R^{13}_4= R^7_{10}= R^{10}_7= R^8_{14}= R^{14}_8= R^{12}_{15}= R^{15}_{12}= e^{-u},\\
&\tau R^2_2  = \tau R^3_3 = \tau R^5_5  =\tau R^9_9  = \kappa  R^8_8= \kappa  R^{12}_{12}= \kappa  R^{14}_{14}= \kappa  R^{15}_{15}=R^4_4= R^7_7= R^{10}_{10}= R^{13}_{13}= e^{-u} u.
\end{align}

Remarkably, for $\tau=\kappa=1$
\begin{align}
R(u)=\frac{u}{e^{u}}\left(1+\frac{1}{u} P \right),
\end{align}
which is the $R$-matrix of the $SU(4)$ invariant chain (apart from an
irrelevant pre-factor).

\section{Model B2}

We have
\begin{align}
&h=\frac{1}{2}\left(
\begin{array}{cccc}
 0 & 0 & 0 & 0 \\
 0 & 0 & e^{i \phi } & 0 \\
 0 & e^{-i \phi } & 0 & 0 \\
 0 & 0 & 0 & 0 \\
\end{array}
\right),
&&\frac{\ell(u)}{\beta(u)}=\left(
\begin{array}{cccc}
 \cosh (u) & 0 & 0 & 0 \\
 0 & 1 & i e^{i \phi } \sinh (u) & 0 \\
 0 & -i \frac{\sinh (u)}{e^{i \phi }}& -1 & 0 \\
 0 & 0 & 0 & -\cosh (u) \\
\end{array}
\right),
\end{align}
and
\begin{equation}
\beta(u)= \sqrt[4]{\frac{\gamma}{2\gamma  \cosh (2 u)+2}}  
\end{equation}
with $\gamma\ge 0$ being a fixed coupling constant.

This model is equivalent to the Hubbard model, and it was already
  studied in \cite{essler-lindblad-reviewS}, see Section 4.3.1 in that
  work. Our formulas involve a simple reparametrization of the
  coupling, and the additional twist $e^{i\phi}$. These changes are
  easily accomodated into the known $R$-matrix, displayed for example
  in \cite{Hubbard-bookS}.

\section{Model B3}

We have
\begin{align}
&h=\frac{1}{2}\left(
\begin{array}{cccc}
 0 & 0 & 0 & 0 \\
 0 & 0 & e^{i \phi } & 0 \\
 0 & e^{-i \phi } & 0 & 0 \\
 0 & 0 & 0 & 0 \\
\end{array}
\right),
&&\ell=\sqrt{\frac{\gamma }{2}} \left(
\begin{array}{cccc}
 \gamma  & 0 & 0 & 0 \\
 0 & 1 & i (\gamma -1) e^{i \phi } & 0 \\
 0 & -i (\gamma +1) e^{-i \phi } & -1 & 0 \\
 0 & 0 & 0 & \gamma  \\
\end{array}
\right),
\end{align}
with $\gamma\ge 0$ being a coupling constrant.

The non-zero entries of the $R$-matrix are
\begin{align}
\nonumber &R_1^1= R^6_6= R^{11}_{11}= R^{16}_{16}= 1,\\
\nonumber &\frac{R^2_2}{e^{i \phi }}= -e^{i \phi }R^3_3 = -e^{i \phi }R^8_8 = \frac{R^{12}_{12}}{e^{i \phi }}= \frac{i (\gamma +1)}{\gamma +\coth (\alpha (u))},\\
\nonumber & -e^{i \phi } R^5_5= \frac{R^9_9}{e^{i \phi }}= \frac{R^{14}_{14}}{e^{i \phi }}=-e^{i \phi } R^{15}_{15} = \frac{i (\gamma -1)}{\gamma +\coth (\alpha (u))},\\
\nonumber &R^2_5= R^5_2= R^{15}_{12}= R^{12}_{15}= R^8_{14}= R^{14}_8= R^9_3= R^3_9= \frac{1}{\gamma  \sinh (\alpha (u))+\cosh (\alpha (u))},\\
\nonumber & e^{i \phi } R^4_7= -\frac{R^4_{10}}{e^{i \phi }}= \frac{R^{13}_{10}}{e^{i \phi }}= -e^{i \phi }R^{13}_7 = \frac{2 i \left(\gamma ^2-1\right) \sinh (\alpha (u))}{\zeta (u)},\\
\nonumber &-e^{i \phi }R^7_4 = R^{10}_4= \frac{2 i (\gamma +1)^2 e^{i \phi } \sinh (\alpha (u))}{\zeta (u)},\\
\nonumber &e^{i \phi }R^7_{13} = -\frac{R^{10}_{13}}{e^{i \phi }}= \frac{2 i (\gamma -1)^2 \sinh (\alpha (u))}{\zeta (u)},
\\
\nonumber &R^{10}_7= R^7_{10}= \frac{2 \left(\gamma ^2+1\right)}{\zeta (u)},\\
\nonumber &e^{2 i \phi }R^7_7 = \frac{R^{10}_{10}}{e^{2 i \phi }}= -\frac{2 \left(\gamma ^2-1\right) \sinh (\alpha (u)) (\gamma  \cosh (\alpha (u))+\sinh (\alpha (u)))}{\zeta (u)},\\
\nonumber &R^4_4= \frac{2 (\gamma +1)^2 \sinh (\alpha (u)) (\gamma  \cosh (\alpha (u))+\sinh (\alpha (u)))}{\zeta (u)},\\
\nonumber &R^4_{13}= \frac{4 e^{2 \alpha (u)} \left(-\gamma +(\gamma +1) \gamma  e^{2 \alpha (u)}+1\right)}{-(\gamma -1)^3-2 \left(\gamma ^2-1\right) e^{2 \alpha (u)}+(\gamma +1)^3 e^{4 \alpha (u)}},\\
\nonumber &R^{13}_4= -\frac{4 \left((\gamma -1) \gamma +(\gamma +1) e^{2 \alpha (u)}\right)}{(\gamma -1)^3+2 \left(\gamma ^2-1\right) e^{2 \alpha (u)}-(\gamma +1)^3 e^{4 \alpha (u)}},\\
\nonumber &R^{13}_{13}= \frac{2 (\gamma -1)^2 \sinh (\alpha (u)) (\gamma  \cosh (\alpha (u))+\sinh (\alpha (u)))}{\zeta (u)},
\end{align}
where $\alpha(u)=\frac{1}{2} \left(\gamma ^2+1\right) u$ and $\zeta(u)=1-\gamma ^2+\gamma ^3 \sinh (2 \alpha (u))+3 \gamma ^2 \cosh (2 \alpha (u))+3 \gamma  \sinh (2 \alpha (u))+\cosh (2 \alpha (u))$.

\providecommand{\href}[2]{#2}\begingroup\raggedright\endgroup

\end{document}